\documentclass[%
showpacs,
 amsmath,amssymb,
 pre, twocolumn
]{revtex4-1}

\pdfoutput=1
\usepackage{graphicx}
\usepackage{dcolumn}
\usepackage{bm}
\usepackage{amssymb}
\usepackage{epstopdf}
\usepackage[top=0.75in, bottom=0.75in, left=0.5in, right=0.5in]{geometry}
\usepackage{enumitem}
\setlist[enumerate]{itemsep=1mm}

\usepackage[utf8]{inputenc}

\usepackage[font=small,labelfont=bf,format=plain,justification=justified,singlelinecheck=false]{caption} 

\begin{document}


\title{Quantum Simulation of Molecular Collisions in the Time-Dependent Formulation}
\author{Andrew T. Sornborger$^{1}$}
\author{Phillip Stancil$^{2}$}
\author{Michael Geller$^2$}
\affiliation{$^1$Department of Mathematics, University of California, Davis, CA 95616, USA}
\affiliation{$^2$Department of Physics, University of Georgia, Athens, GA 30606, USA}

\date{\today}

\begin{abstract}
\noindent 
Quantum particle simulations have largely been based on time-independent, split-operator schemes in which kinetic and potential operators are interwoven to provide accurate approximations to system dynamics. These simulations can be very expensive in terms of the number of gates required, although individual cases, such as tunneling, have been found where prethreshold simulations are possible. By prethreshold, we mean a quantum computation or simulation with an existing architecture and not requiring error correction. In the case of molecular collisions, switching to a time-dependent formulation can reduce the system dimensionality significantly and provide an opportunity for pre-threshold simulation. Here, we study the efficiency of gate-based quantum simulation of a set of molecular collisions of increasing complexity. We show that prethreshold quantum simulation of such systems is feasible up to Hilbert space dimension 8, but beyond that error correction would be required.
\end{abstract}


\maketitle


\section{Introduction}

Quantum simulations are one of a set of quantum algorithms that give exponential improvement relative to the best classical algorithm. Small quantum simulations have been realized on NMR \citep{TsengEtAl1999,SomarooEtAl1999,KhitrinFung2001,NegrevergneEtAl2005,PengEtAl2005,BrownEtAl2006,PengEtAl2009,DuEtAl2010}, atomic \citep{EdwardsEtAl2010,KinoshitaEtAl2004}, ion trap \citep{FriedenauerEtAl2008,GerritsmaEtAl2010,GerritsmaEtAl2011,LanyonEtAl2011,LanyonEtAl2010} and photonic \citep{MaEtAl2011,KassalEtAl2011} quantum computers in two forms: analog simulations in which a many-body or multiple spin Hamiltonian is mapped to a computational Hamiltonian, and digital simulations in which a Hamiltonian is split into free and interacting operators, then is simulated on a quantum computer. Digital quantum particle simulations \citep{Zalka1998} such as those proposed for the simulation of chemical dynamics \citep{KassalEtAl2008} remained untested until recently \cite{Sornborger2012,FengEtAl2013} due to the large number of gates and/or ancillary qubits needed to compute the kinetic and potential operators \citep{KassalEtAl2008,BenentiStrini2008}. 

In previous work we have explored a simplified gate-based approach to quantum particle simulation that results in short enough circuits to avoid the need for error correction \cite{Sornborger2012}. We refer to any quantum computation or simulation that uses existing hardware and has a short enough circuit depth requirement to avoid error correction as {\it prethreshold}, referring to the threshold theorems of fault tolerant quantum computation. We have also shown how prethreshold universal quantum simulation can be implemented with a complete graph of superconducting qubits operating in the single-excitation subspace \cite{GellerEtAl2015}. In particular, we showed how to solve time-dependent molecular collision dynamics problems with a single-excitation subspace quantum simulator.

In this paper, we will investigate the difficulty, and possibility for pre-threshold simulation, of encoding a time-dependent formulation of molecular collision dynamics on a gate-based quantum computer. To do this, we use an arsenal of both Magnus and operator splitting methods to approximate the evolution of scattering amplitudes due to a time-dependent Hamiltonian on a gate-based quantum computer. 

In Section \ref{sec2}, we describe a means of mapping an arbitrary, time-dependent Hamiltonian to a target quantum Hamiltonian for a quantum computer of interest. We then catalog the set of unitary methods that we use for approximating evolution induced by a time-dependent Hamiltonian. In Section \ref{sec3}, we present results of simulations of a set of molecular collisions of increasing complexity. These results make clear the advantages and disadvantages of the approximation methods that we use. Finally, in Section \ref{sec4}, we discuss our results and their implications in terms of the best methods for various simulational needs.

\section{Methods}\label{sec2}

\subsection*{Molecular Collisions}

We use the semiclassical formulation of a molecular collision \cite{Child1984}. Scatterers are assumed to follow a straight-line trajectory. For an impact at closest approach occurring at time $t_0$, the internuclear separation varies according to $$R = \sqrt{b^2 + v_0^2 (t - t_0)^2} \; ,$$ where $v_0$ is the initial relative velocity and $b$ the impact parameter of the collision. The relative velocity is related to the collision energy in the center-of-mass frame through $E_{\rm{cm}} = \mu v_o^2/2$, where $\mu$ is the collision system's reduced mass.

We construct an $n \times n$ potential-coupling matrix $U(R)$. The diagonal elements of $U$ are diabatic electronic potential energies constructed using molecular state bases. The scattering Hamiltonian $H(R)$ is defined with diagonal elements $$H_{ii}(R) = U_{ii}(R) + \frac{\mu}{2} \left( \frac{bv_0}{R} \right)^2 \; .$$
The off-diagonal elements of $U(R)$ are nonadiabatic couplings expressed in a diabatic representation. Three different collision complexes with increasing number of channels $n$ are considered:
Na + He excitation \cite{LinEtAl2008}, Si$^{3+}$ + He charge exchange \cite{StancilEtAl1999}, and O$^{7+}$ + H charge exchange \cite{NolteEtAl2016}.

In order to translate the physical Hamiltonian to a quantum computational Hamiltonian \cite{GellerEtAl2015}, we first subtract an overall (unmeasurable) time-dependent phase 
$$c(t) \equiv \frac{1}{n} \sum_{i=1}^N H_{ii} \; .$$ 
We then find, at each time $t$, the smallest positive $\lambda$ such that every matrix element of 
$$\mathcal{H}(t) = \frac{H(t) - c(t) \times I}{\lambda(t)}$$ 
is between $-g_{max}$ and $g_{max}$. The simulated time is then $$t_{\rm{qc}} = \int_0^t \lambda(t') dt' \; .$$ The function $t_{\rm{qc}}$ is then inverted and the remapped quantum evolution is described by $$\dot{\psi} = -i \mathcal{H}(t_{\rm{qc}}) \psi \; .$$ This rescaling, a continuous form of variable time slicing, maps the physical Hamiltonian to the quantum simulation energy scale. It additionally assures that the quantum simulation is performed as efficiently as possible by rescaling time such that the simulation speeds through parts of the collision where not much is happening, but slows down when the collision dynamics must be accurately integrated. 

\subsection*{Unitary Operator Approximation Methods}\label{sec2}

For ease of notation, we will now write the Schr\"odinger equation as
\begin{equation}
\dot{\psi} = S(t) \psi \; ,\label{Sch} \nonumber
\end{equation}
where $S(t) = -i \mathcal{H}(t)$ is skew-Hermitian.

The time evolution of the wavefunction, $\psi(t)$, is given by a time-ordered matrix exponential, typically written as
\begin{equation}
\psi(t) = \mathcal{T}\left\{ e^{\int_0^t S(t') dt'} \right\} \psi(0) \; ,\nonumber
\end{equation}
where the time-ordered matrix exponential may be expanded \cite{Magnus1954} as
\begin{widetext}
\begin{eqnarray}
\mathcal{T}\left\{ e^{\int_0^t S(t') dt'} \right\} & = & \exp\biggl( \int_0^t S(\tau) d\tau + \frac{1}{2} \int_0^t \left[ S(\tau), \int_0^\tau S(\tau') d\tau' \right] d\tau \nonumber \\
& & + \frac{1}{4} \int_0^t \left[ S(\tau), \int_0^\tau \left[ S(\tau'), \int_0^{\tau'} S(\tau'') d\tau'' \right] d\tau' \right] d\tau \nonumber \\
& & + \dots \biggl) \; .\nonumber
\end{eqnarray}
\end{widetext}
The crudest simulation of this evolution may be done by fixing the time step, $\Delta t$, then, assuming $S(t)$ to be slowly varying during each timestep. In this case, we have
\begin{equation}
\prod_{n = 0}^N e^{S(n \Delta t) \Delta t} = \mathcal{T}\left\{ \exp \left( \int_0^t S(t') dt' + O(N \Delta t^2) \right) \right\}\; .
\end{equation}
Here, $N = T/\Delta t$, with $T$ the total simulation time. In this matrix product (and below), unitaries are put in decreasing order of $n$, thus early-time operators act first on the wavefunction, $\psi$. In the Results section, we will refer to this first-order in time, unsplit scheme as a $(1,0,0)$ scheme. Here, and below, the notation to denote each approximation scheme is $(a, b, c)$, where $a$ denotes the order of the Magnus-splitting, $b$ denotes the order of the split-operator approximation of each operator within the Magnus-splitting (see below), and $c$ denotes the order of the split-operator approximation of three-qubit operators via commutators of two-qubit operators (see below).

We can achieve higher accuracy than that in the $(1,0,0)$ scheme. Defining
\begin{equation}
S_1 = S \left(t_n + \left(\frac{1}{2} - \frac{\sqrt{3}}{6}\right) \Delta t \right) \nonumber
\end{equation}
and
\begin{equation}
S_2 = S \left(t_n + \left(\frac{1}{2} + \frac{\sqrt{3}}{6}\right) \Delta t \right) \; ,\nonumber
\end{equation}
the time-ordered matrix exponential may be approximated \cite{IserlesEtAl1999} as
\begin{widetext}
\begin{equation}
\exp\left( S_n (\Delta t) \right) \equiv \exp\left( \frac{1}{2} \Delta t (S_1 + S_2) - \frac{\sqrt{3}}{12} \Delta t^2 \left[ S_1, S_2 \right] \right) = \mathcal{T}\left\{ \exp \left( \int_{t_n}^{t_n + \Delta t} S(t') dt' + O(\Delta t^5) \right)  \right\} \; .\nonumber
\end{equation}
\end{widetext}
Providing $S(t)$ is continuous and non-singular \cite{WiebeEtAl2010}, this gives a fourth-order accurate unitary approximation for a single time-step. To approximate the full evolution, we form the product
\begin{equation}
\prod_{n = 0}^{N} \exp\left( S_n  (\Delta t) \right) = \mathcal{T}\left\{ \exp \left( \int_0^t S(t') dt' + O(N \Delta t^5) \right) \right\} \; ,\nonumber
\end{equation}
giving a fourth-order accurate time-stepping scheme. Here, $S_n$ is evaluated at equal intervals on $[0, t]$. We will refer to this as a $(4,0,0)$ scheme.

To implement these methods in a quantum simulation, we must consider the capabilities of the quantum computer. The time-independent unitary evolution, $\exp(S_n (\Delta t))$, of a single step may or may not be able to be implemented in one gate. We will assume that our schemes are implemented on a quantum computer capable of forming only single- and two-qubit operations. Thus, simulations on one- or two-qubits may be implemented directly, but three-qubit simulations will require further modifications. 

In order to deal with these issues, we first expand in a basis of orthogonal matrices, $\{ \gamma_j \}$,
\begin{equation}
S_n = \sum_j a^n_j \gamma_j \; ,\nonumber
\end{equation}
where $j \in 1, \dots, J$, with $J$ the number of basis matrices, and $a^n_j = \langle \gamma_j, S_n \rangle$, assuming $\gamma_j$ is unit norm.

Assuming that each $\exp(a^n_j \gamma_j \Delta t)$ may be implemented on our quantum computer, a first-order split operator approximation may be made to $\exp(S_n (\Delta t))$:
\begin{equation}
\prod_j \exp(a^n_j \gamma_j \Delta t) =  \exp\left(S_n \Delta t + O(\Delta t^2) \right) \; . \nonumber
\end{equation}
We will refer to this as a $(4,1,0)$ scheme.

Using the notation
$\prod_j \exp(a^n_j \gamma_j m \Delta t) \equiv (m \Delta t)_n$,
we can write the full evolution approximated by this and higher-order schemes \cite{SornborgerStewart1999} as
\begin{widetext}
\begin{subequations}\label{splitmeth}
\begin{eqnarray}
\mathcal{T} \left\{ \exp\left(\int_0^t S(t') dt' + O(N\Delta t^2) \right) \right\} & = & \prod_n (\Delta t)_n \label{splitmeth1} \\
\mathcal{T} \left\{ \exp\left(\int_0^t S(t') dt' + O(N\Delta t^3) \right) \right\} & = & \prod_n \left(\frac{\Delta t}{2} \right)_n \left(\frac{\Delta t}{2}\right)_n^T \label{splitmeth2} \\
\mathcal{T} \left\{ \exp\left(\int_0^t S(t') dt' + O(N\Delta t^5) \right) \right\} & = & \prod_n \left(\frac{\Delta t}{12}\right)_n^T \left(\frac{\Delta t}{12}\right)_n \left(\frac{\Delta t}{12}\right)_n^T \left(\frac{-\Delta t}{6}\right)_n \left[ \left(\frac{\Delta t}{12}\right)_n^T \right]^4 \left(\frac{\Delta t}{12}\right)_n \nonumber \\ & \; & \;\;\; \left(\frac{\Delta t}{12}\right)_n^T \left[ \left(\frac{\Delta t}{12}\right)_n \right]^4 \left(\frac{-\Delta t}{6}\right)_n^T \left(\frac{\Delta t}{12}\right)_n \left(\frac{\Delta t}{12}\right)_n^T \left(\frac{\Delta t}{12}\right)_n \label{splitmeth3} \; ,
\end{eqnarray}
\end{subequations}
\end{widetext}
which we will refer to as $(4,1,0)$ (as above), $(4,2,0)$, and $(4,4,0)$ schemes, respectively. The $^T$ denotes the transpose here, not the Hermitian transpose. Note that these schemes use more operators to achieve higher-order accuracy.

\begin{figure*}[t]
\includegraphics[width=\textwidth]{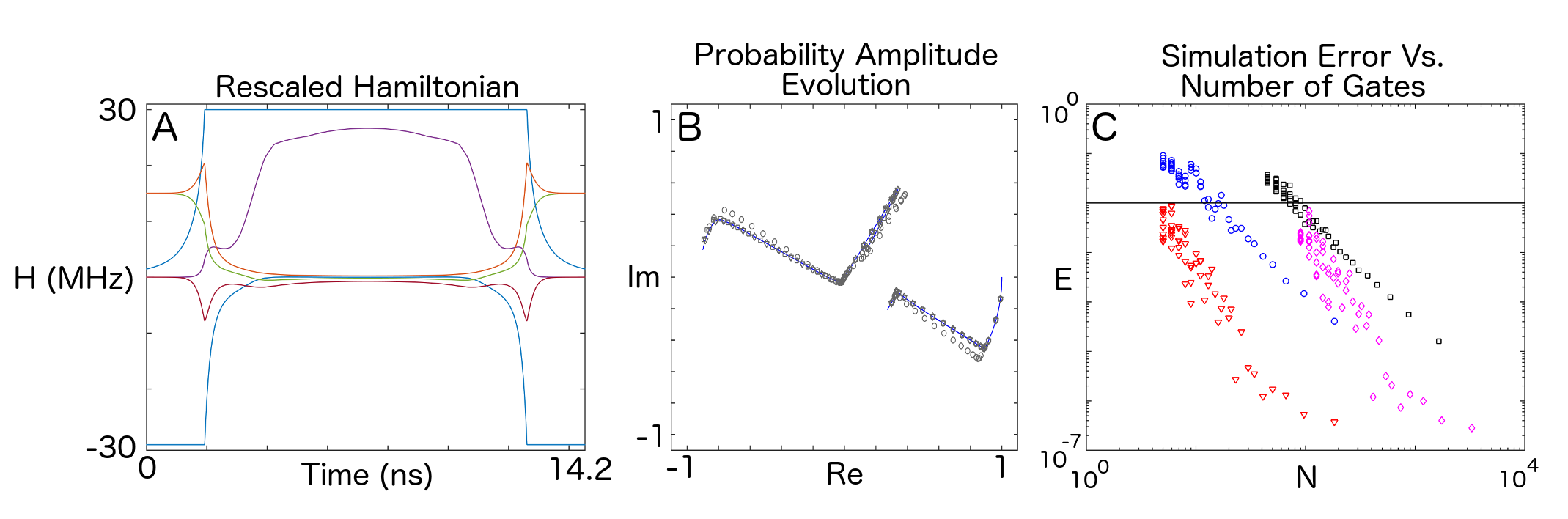}
\caption{Na + He Collision, 3 Channels. $v_0 = 2$ a.u., $b = 0.5$ a.u.. A) Rescaled Hamiltonian. Here, we assume that $g_{\rm{max}}/2\pi = 30$ MHZ (typical for a superconducting quantum computer). At each instant the magnitude of at least one matrix element achieves its maximum value of $30$ MHZ, making the simulation as fast as possible. B) The time evolution of the probability amplitude for a simulation with $N = 21$. The initial condition is that the first channel's amplitude is $1$ and the rest are set to $0$, thus one branch in this panel starts at $1$, but the other two branches start at $0$. Blue lines indicate the exact evolution, with a $(1,0,0)$ simulation given by gray circles, $(4,0,0)$ - triangles, $(4,1,0)$ - squares, and $(4,2,0)$ - diamonds. Note that for this number of timesteps, all Magnus method simulations give fairly accurate results. C) Simulation error plotted as a function of the number of gates, with $(1,0,0)$ simulation given by blue circles, $(4,0,0)$ - red triangles, $(4,1,0)$ - black squares, and $(4,2,0)$ - magenta diamonds.}
\end{figure*}

The above schemes work perfectly well for simulations that may be encoded in one- and two-qubits. However, once three-qubit operators, $\{ \gamma^3_j \}$, are needed for the simulation, they must somehow be generated. This may be done using approximations to exponentiated commutators, $\exp([A,B]\Delta t)$. 

For instance, the commutator between arbitrary three-qubit operators takes the form
\begin{widetext}
$$\left[ \sigma_i \otimes \sigma_j \otimes \sigma_k, \sigma_a \otimes \sigma_b \otimes \sigma_c \right] = \sigma_i \sigma_a \otimes  \sigma_j \sigma_b \otimes  \sigma_k \sigma_c - \sigma_a \sigma_i \otimes  \sigma_b \sigma_j \otimes  \sigma_c \sigma_k \; ,$$
\end{widetext}
where the $\sigma$'s are single-qubit operators (i.e. Pauli matrices). So, for example, if we take $\sigma_i = \sigma_0 = I_2$ and $\sigma_c = \sigma_0 = I_2$, where $I_2$ is the $2 \times 2$ identity matrix, then the commutator between the two-qubit operators 
\begin{widetext}
$$\left[ \gamma^2_{jk}, \gamma^2_{ab} \right] = \left[ I_2 \otimes \sigma_j \otimes \sigma_k, \sigma_a \otimes \sigma_b \otimes I_2 \right] = \sigma_a \otimes  \sigma_j \sigma_b \otimes  \sigma_k - \sigma_a \otimes  \sigma_b \sigma_j \otimes  \sigma_k = \sigma_f \otimes \sigma_g \otimes \sigma_h \; ,$$
\end{widetext}
with $\sigma_f \otimes \sigma_g \otimes \sigma_h$ a three-qubit operator. We have checked that all three-qubit operators can be generated from two-qubit operators in this way.

We use two approximations of exponentiated commutators, one with $O(\Delta t^{3/2})$ error \cite{Lloyd1995} and one with $O(\Delta t^{5/2})$ error \cite{SornborgerStewart1999}:
\begin{widetext}
\begin{eqnarray}
\exp\left(a \gamma^3 \Delta t + O(\Delta t^{3/2})\right) & = & \exp\left(a [\gamma_i^2, \gamma_j^2] \Delta t) + O(\Delta t^{3/2}) \right) \nonumber \\
& = & \exp(-\gamma_i^2 \sqrt{-a\Delta t}) \exp(- \gamma_j^2 \sqrt{-a\Delta t}) \exp( \gamma_i^2 \sqrt{-a\Delta t}) \exp(\gamma_j^2 \sqrt{-a\Delta t}) \label{commute1} \\
& \equiv & \left( -\sqrt{-\Delta t} \right) \left( \sqrt{-\Delta t} \right) \nonumber \\
\exp\left(a \gamma^3 \Delta t + O(\Delta t^{5/2})\right) & = & \left( -2\sqrt{\frac{\Delta t}{12}} \right)^T \left( 2\sqrt{\frac{\Delta t}{12}} \right)^T \left[ \left(-\sqrt{\frac{\Delta t}{12}}\right) \left(\sqrt{\frac{\Delta t}{12}}\right) \right]^{12} \left[ \left(\sqrt{\frac{\Delta t}{12}} \right) \left(-\sqrt{\frac{\Delta t}{12}} \right) \right]^4 \label{commute2} \; ,
\end{eqnarray}
\end{widetext}
where $\gamma^2_j$ denotes a two-qubit basis matrix.

When these approximations are used to replace the three-qubit unitaries in (\ref{splitmeth}), we refer to them as $(4,1,\frac{3}{2})$, $(4,2,\frac{3}{2})$, $(4,4,\frac{3}{2})$, and $(4,1,\frac{5}{2})$, $(4,2,\frac{5}{2})$, and $(4,4,\frac{5}{2})$ schemes. Note the significant cost in operations to approximate three-qubit operations with two-qubit operations.

\begin{figure*}[t]
\includegraphics[width=\textwidth]{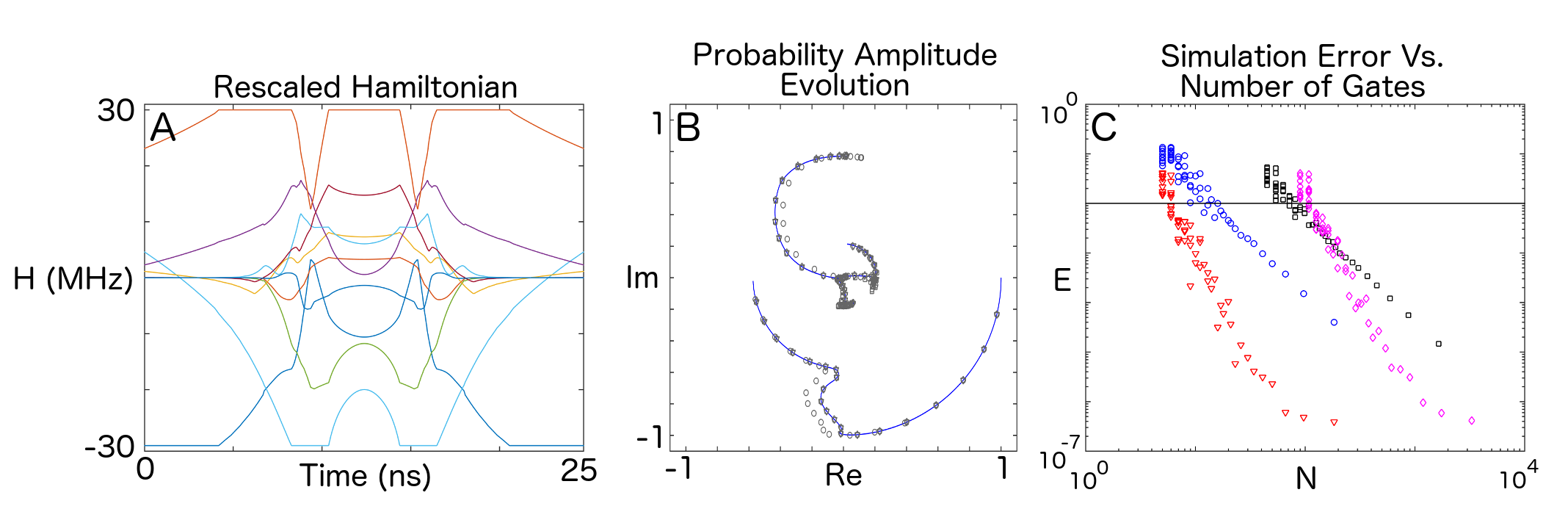}
\caption{Si$^{3+}$ + He Collision, 4 Channels. $v_0 = 2$ a.u., $b = 0.5$ a.u.. A) Rescaled Hamiltonian. Here, we assume that $g_{\rm{max}}/2\pi = 30$ MHZ. B) The time evolution of the probability amplitude for a simulation with $N = 21$. The initial condition is that the first channel's amplitude is $1$ and the rest are set to $0$, thus one branch in this panel starts at $1$, but the other branches start at $0$. Blue lines indicate the exact evolution, with a $(1,0,0)$ simulation given by circles, $(4,0,0)$ - triangles, $(4,1,0)$ - squares, and $(4,2,0)$ - diamonds. C) Simulation error plotted as a function of the number of gates, with $(1,0,0)$ simulation given by blue circles, $(4,0,0)$ - red triangles, $(4,1,0)$ - black squares, and $(4,2,0)$ - magenta diamonds.}
\end{figure*}

\section{Results}\label{sec3}

Our goal is to determine whether gate-based quantum simulations of molecular collisions in the time-dependent formulation is feasible on a pre-threshold quantum computer with the methods described above. We investigate three collision systems which have increasing numbers of channels and internal energy scales. The first involves the electronic excitation of Na due to He simulated with three channels with off-diagonal
radial and rotational nonadiabatic couplings \cite{LinEtAl2008}. Two charge exchange reactions are considered including Si$^{3+}$ + He in both 4- and 5-channel simulations and O$^{7+}$ + H with eight channels
\cite{StancilEtAl1999,NolteEtAl2016}. As charge exchange is dominated by radial coupling, the rotational nonadiabatic couplings are neglected. All systems are treated in a diabatic representation.

\subsection{Two-Qubit Simulations}

{\it Na + He Collision - 3-channel} A 3-channel Sodium Helium collision simulation may be encoded in two qubits. To do this, we placed the $3 \times 3$ quantum collision Hamiltonian in the first three rows and columns of a $4 \times 4$ su(4) operator. Note that the collision Hamiltonians, here and below, whose dynamics we simulate had all real elements, requiring fewer parameters than an arbitrary matrix in su(4). We investigated four quantum simulation methods: $(1,0,0)$, $(4,0,0)$, $(4,1,0)$, and $(4,2,0)$. In Fig. 1, we plot the rescaled Hamiltonian (A), the evolution of the probability amplitudes in the complex plane (B), and the logarithm of the number of gates required versus the logarithm of the error for each simulation (C). Errors were defined as $$E = \left\langle \left| \psi_j(t_{\rm{final}})_{\rm{sim}} - \psi_j(t_{\rm{final}})_{\rm{exact}} \right|^2 \right\rangle_j \; ,$$ where $\langle \rangle_j$ denotes an average over all states. In this case, $j = 1, \dots, 4$. For `exact' simulations, we computed the evolution on a fine temporal grid on which the simulation had converged. To compare to an exact simulation, we performed simulations on a range of temporal grids with timestep $\Delta t$, giving $N = T/\Delta t$ timesteps.

The main points to notice here (Fig. 1C) are that the simulation using the $4$'th-order Magnus method, $(4,0,0)$, gave a much smaller error, $E$, for a given number of gates, $N$, than the first-order, `crude' method, $(1,0,0)$. Additionally, once the simulations were split into single- and two-qubit gates, $(4,1,0)$ and $(4,2,0)$, there was a significant increase in the number of gates required to perform the simulation and the overall slope of the dots that represent each simulation in the figure became somewhat less negative due to errors arising from the splitting. Setting a minimum of $99\%$ fidelity (i.e. $E < 0.01$) for an accurate simulation, we found that the $(1,0,0)$ method required at least $13$ arbitrary two-qubit gates, the $(4,0,0)$ method required $5$ arbitrary two-qubit gates, the $(4,1,0)$ method required $32$ single-qubit and $40$ two-qubit gates, and the $(4,2,0)$ method required $40$ single-qubit and $50$ two-qubit gates.

{\it Si$^{3+}$ + He Collision - 4-channel} This simulation required no embedding and the 4 channel Hamiltonian was mapped directly into an su(4) operator. Again, we investigated four methods: $(1,0,0)$, $(4,0,0)$, $(4,1,0)$, and $(4,2,0)$. In Fig. 2C, we see that the Magnus method still outperformed the crude method, and we again see the extra gates needed once the operators are split. This simulation (and subsequent simulations) required more computational time due to the more complicated dynamics represented in the collision Hamiltonian (Fig. 2A). For $99\%$ fidelity, we found that the $(1,0,0)$ method required at least $12$ arbitrary two-qubit gates, the $(4,0,0)$ method required $6$ arbitrary two-qubit gates, the $(4,1,0)$ methods required $32$ single- and $40$ two-qubit gates, and the $(4,2,0)$ method required $48$ single-qubit and $60$ two-qubit gates.

\subsection{Three-Qubit Simulations}

\begin{figure*}[t]
\includegraphics[width=\textwidth]{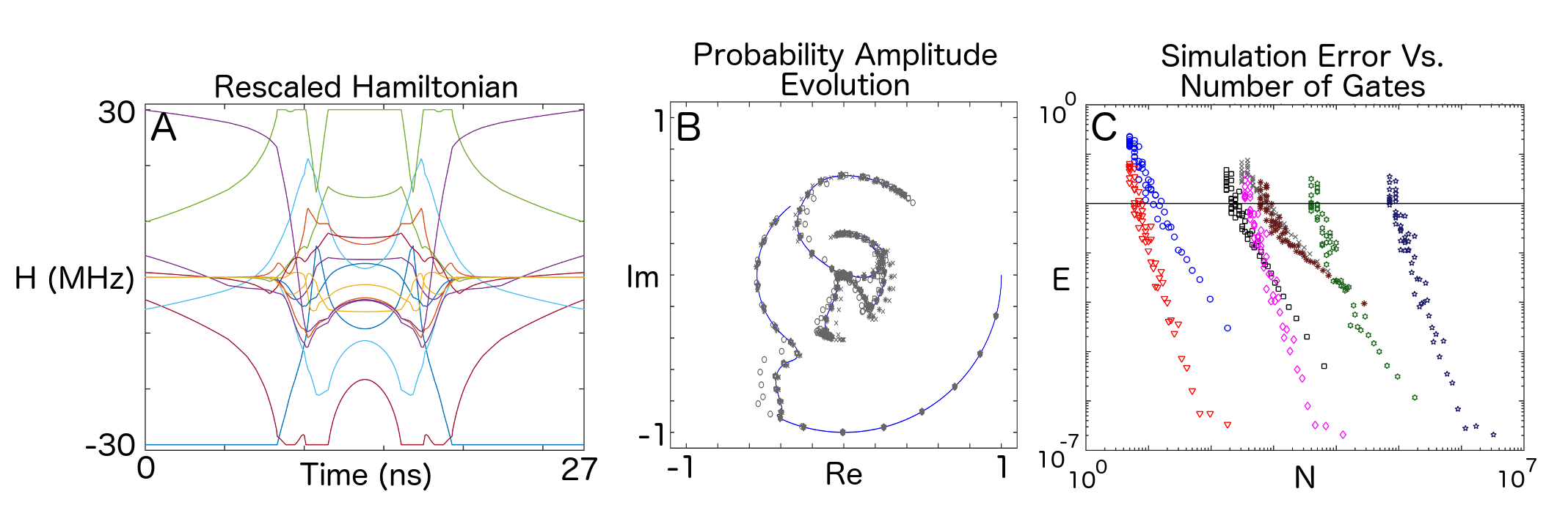}
\caption{Si$^{3+}$ + He Collision, 5 Channels. $v_0 = 2$ a.u., $b = 0.5$ a.u.. A) Rescaled Hamiltonian. Here, we assume that $g_{\rm{max}}/2\pi = 30$ MHZ. B) The time evolution of the probability amplitude for a simulation with $N = 21$. The initial condition is that the first channel's amplitude is $1$ and the rest are set to $0$, thus one branch in this panel starts at $1$, but the other branches start at $0$. Blue lines indicate the exact evolution, with a $(1,0,0)$ simulation given by circles, $(4,0,0)$ - triangles, $(4,1,0)$ - squares, and $(4,2,0)$ - diamonds, $(4,1,\frac{3}{2})$ - x's, $(4,2,\frac{3}{2})$ - *'s, $(4,2,\frac{5}{2})$ - hexagrams, and $(4,4,\frac{5}{2})$ - pentagrams. C) Simulation error plotted as a function of the number of gates, with $(1,0,0)$ simulation given by blue circles, $(4,0,0)$ - red triangles, $(4,1,0)$ - black squares, and $(4,2,0)$ - magenta diamonds, $(4,1,\frac{3}{2})$ - gray x's, $(4,2,\frac{3}{2})$ - reddish brown *'s, $(4,2,\frac{5}{2})$ - green hexagrams, and $(4,4,\frac{5}{2})$ - dark blue pentagrams. We include the $(1,0,0)$, $(4,0,0)$, $(4,1,0)$, and $(4,2,0)$ simulations for comparison with the two-qubit results. Note that, for a two-qubit gate quantum computer, they are unphysical.}
\end{figure*}

{\it Si$^{3+}$ + He Collision - 5-channel} The 5-channel Silicon-Helium collision simulation must be embedded in su(8) and requires three qubits. We placed the $5 \times 5$ collision Hamiltonian in the first five rows and columns of an $8 \times 8$ su(8) operator. We investigated the methods $(1,0,0)$, $(4,0,0)$, $(4,1,0)$, $(4,2,0)$, $(4,1,\frac{3}{2})$, $(4,2,\frac{3}{2})$, $(4,2,\frac{5}{2})$ and $(4,4,\frac{5}{2})$. Note, that although we show results for them, now the $(1,0,0)$, $(4,0,0)$, $(4,1,0)$, and $(4,2,0)$ methods are unphysical, since they would require three-qubit operators. In Fig. 3C, we see that, in addition to the additional gates needed to split the su(8) operator into single-, two-, and three-qubit gates, yet more gates were needed to convert three-qubit gates to two-qubits via the exponentiated commutator approximation methods, (\ref{commute1}) and (\ref{commute2}).
\begin{widetext}
\begin{minipage}{\linewidth}
\centering
\captionof{table}{Si$^{3+}$ + He 5-Channel Collision Simulation Results} \label{tab1} 
\begin{tabular}{| l | l | l | l | l | l | l | l | l |}
\hline
Method & $(1,0,0)$ & $(4,0,0)$ & $(4,1,0)$ & $(4,2,0)$ & $(4,1,\frac{3}{2})$ & $(4,2,\frac{3}{2})$ & $(4,2,\frac{5}{2})$ & $(4,4,\frac{5}{2})$ \\
\hline
1-qubit & 0   & 0 & 36   & 72   & 66   & 48   & 48     & 864 \\
\hline
2-qubit & 12 & 6 & 174 & 348 & 781 & 568 & 3928 & 70704 \\
\hline
\end{tabular}
\bigskip
\caption*{Number of single- and two-qubit gates needed for $99\%$ fidelity of the final (scattered) wavefunction for the various approximation methods.}
\end{minipage}
\end{widetext}
In Table \ref{tab1}, we see that the Magnus method that used second-order splitting and $\frac{3}{2}$-order commutator approximation could be implemented with $48$ single-qubit operators and $568$ two-qubit operators. This was the best case implementation of this molecular collision Hamiltonian on a standard three-qubit quantum computer that uses two-qubit operators.

\begin{figure*}[t]
\includegraphics[width=\textwidth]{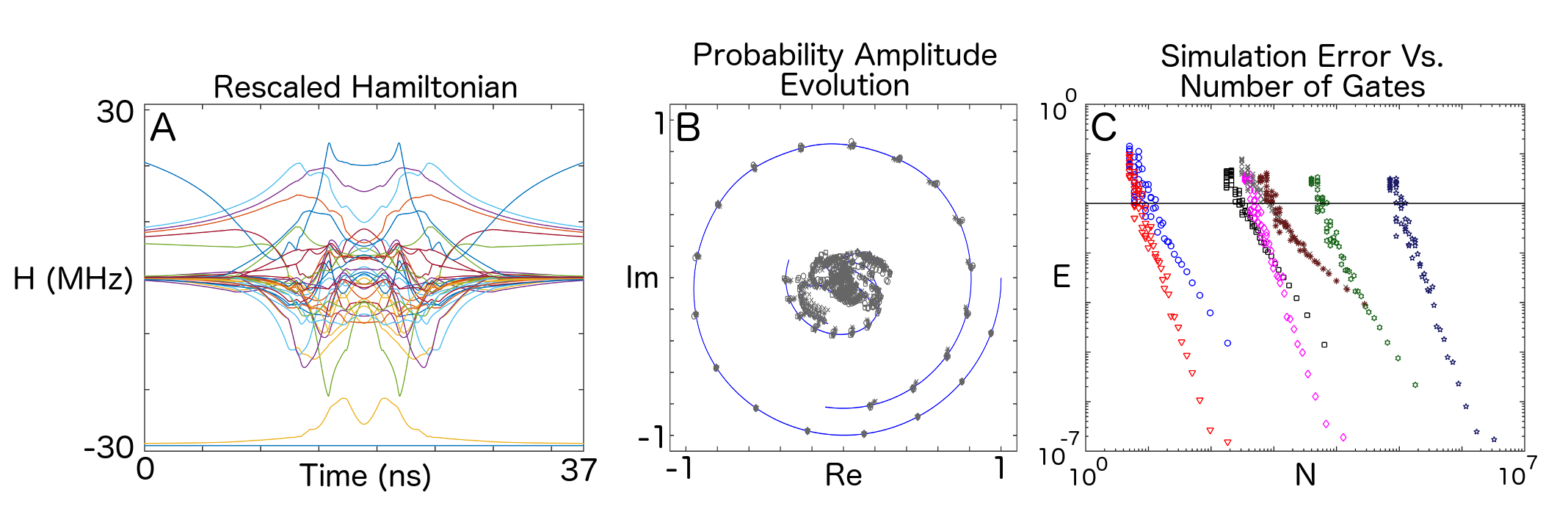}
\caption{O$^{7+}$ + H Collision, 8 Channels. $v_0 = 2$ a.u., $b = 0.5$ a.u.. A) Rescaled Hamiltonian. Here, we assume that $g_{\rm{max}}/2\pi = 30$ MHZ. B) The time evolution of the probability amplitude for a simulation with $N = 21$. The initial condition is that the first channel's amplitude is $1$ and the rest are set to $0$, thus one branch in this panel starts at $1$, but the other branches start at $0$. Blue lines indicate the exact evolution, with a $(1,0,0)$ simulation given by circles, $(4,0,0)$ - triangles, $(4,1,0)$ - squares, and $(4,2,0)$ - diamonds, $(4,1,\frac{3}{2})$ - x's, $(4,2,\frac{3}{2})$ - *'s, $(4,2,\frac{5}{2})$ - hexagrams, and $(4,4,\frac{5}{2})$ - pentagrams. C) Simulation error plotted as a function of the number of gates, with $(1,0,0)$ simulation given by blue circles, $(4,0,0)$ - red triangles, $(4,1,0)$ - black squares, and $(4,2,0)$ - magenta diamonds, $(4,1,\frac{3}{2})$ - gray x's, $(4,2,\frac{3}{2})$ - reddish brown *'s, $(4,2,\frac{5}{2})$ - green hexagrams, and $(4,4,\frac{5}{2})$ - dark blue pentagrams. We include the $(1,0,0)$, $(4,0,0)$, $(4,1,0)$, and $(4,2,0)$ simulations for comparison with the two-qubit results. Note that, for a two-qubit gate quantum computer, they are unphysical.}
\end{figure*}

{\it O$^{7+}$ + H Collision - 8-channel} The 8-channel Oxygen-Hydrogen collision simulation may be encoded directly in su(8). We investigated the same methods as for the Silicon-Helium collision simulations.
\begin{widetext}
\begin{minipage}{\linewidth}
\centering
\captionof{table}{O$^{7+}$ + H 8-Channel Collision Simulation Results} \label{tab2} 
\begin{tabular}{| l | l | l | l | l | l | l | l | l |}
\hline
Method & $(1,0,0)$ & $(4,0,0)$ & $(4,1,0)$ & $(4,2,0)$ & $(4,1,\frac{3}{2})$ & $(4,2,\frac{3}{2})$ & $(4,2,\frac{5}{2})$ & $(4,4,\frac{5}{2})$ \\
\hline
1-qubit & 0 & 0 & 48   & 72   & 66   & 84   & 60     & 1080 \\
\hline
2-qubit & 9 & 6 & 232 & 348 & 781 & 994 & 4910 & 88380 \\
\hline
\end{tabular}
\bigskip
\caption*{Number of single- and two-qubit gates needed for $99\%$ fidelity of the final (scattered) state for the various approximation methods.}
\end{minipage}
\end{widetext}
In Table \ref{tab2}, as in the 5-channel Silicon-Helium collision simulations, the best case implementation of the Oxygen-Hydrogen collision simulations was the $(4,2,\frac{3}{2})$ method using $84$ single-qubit operators and $994$ two-qubit operators.

\section{Discussion}\label{sec4}

We have shown systematic reductions of a set of time-dependent, molecular collision Hamiltonians for simulation on a quantum computer. Via approximation methods of differing accuracy and numbers of gates, we showed the range of difficulties of implementing the resulting quantum simulations. The main conclusion of this work is that molecular collision simulations that may be encoded in two qubits are clearly feasible for implementation on a pre-threshold quantum computer, requiring fewer than $100$ single- and two-qubit operators. Extending such simulations to three qubits will be significantly more challenging, requiring about an order of magnitude more two-qubit gates. The additional gates arise due to the need to approximate exponentiated commutators that generate three-qubit operators from two-qubit operators. These unitaries are expensive in terms of the number of gates required.

We think it will be difficult to further reduce the number of gates that we used in our simulations. Both the Iserles, et al.'s \cite{IserlesEtAl1999} Magnus-method and the high-order operator splitting methods \cite{SornborgerStewart1999} that we used are the most parsimonious methods at 4'th-order that we are aware of in the literature. One possible further avenue for gate reduction may lie in the fact that the size of the elements of $\mathcal{H}(t)$ decreases fairly rapidly as a function of distance from the diagonal, thus at some point a truncation of terms might be possible, resulting in a banded operator that could potentially be simulated with a non-exponentially increasing number of operators. If this is not possible, then the number of Hamiltonian elements necessary to implement a simulation will increase exponentially with the number of channels simulated, making this approach infeasible. As it stands, we have put forth methods for simulating $2^n$ channels with $n$ qubits, giving an exponential savings in system size. Although this savings is trivial for the system sizes we simulate, it may be some time before error-corrected simulations are possible and meaningful statements can be made for large, gate-based quantum simulations.

\begin{acknowledgments}
This work was supported by the National Science Foundation under CDI grant DMR-1029764.
\end{acknowledgments}

\bibliographystyle{unsrtnat}
\bibliography{myrefsv2}

\end{document}